# Network Analyzer Measurements of Spin Transfer Torques in Magnetic Tunnel Junctions

Lin Xue[1], Chen Wang[1], Yong-Tao Cui[1], J. A. Katine[2], R. A. Buhrman[1] and D. C. Ralph[1,3]

[1]Cornell University, Ithaca, New York 14853, USA

[2]Hitachi Global Storage Technologies, San Jose, California 95135, USA

[3]Kavli Institute at Cornell, Ithaca, New York 14853, USA

**Abstract**

We demonstrate a simple network-analyzer technique to make quantitative measurements of the bias dependence of spin torque in a magnetic tunnel junction. We apply a microwave current to exert an oscillating spin torque near the ferromagnetic resonance frequency of the tunnel junction's free layer. This produces an oscillating resistance that, together with an applied direct current, generates a microwave signal that we measure with the network analyzer. An analysis of the resonant response yields the strength and direction of the spin torque at non-zero bias. We compare to measurements of the spin torque vector by time-domain spin-torque ferromagnetic resonance.

Spin transfer torque provides the possibility of efficiently manipulating the magnetic moment in a nanoscale magnetic device using applied current.[1-3] Understanding the strength of the spin torque, and particularly its bias dependence, is important for applications that include spin torque magnetic random access memory and frequency-tunable microwave oscillators.[4] Several different techniques have been developed to measure the bias dependence of the spin torque vector in magnetic tunnel junctions (MTJs), with results that in some cases are inconsistent with each other. These include measurements of the bias dependence of the magnetic precession frequency and linewidth,[5-10] DC-voltage-detected spin torque ferromagnetic resonance (ST-FMR),[11-13] fits to the statistics of magnetic switching as a function of current and magnetic field,[14,15] analyses of the current dependence of magnetic astroids and switching phase diagrams,[16,17] and time-domain detection of ST-FMR.[18] Of these, in the high bias regime that is relevant for applications we believe that the time-domain ST-FMR technique is the most accurate and trustworthy, since it measures directly the amplitude and phase of small-angle magnetic precession in response to an oscillating spin torque and therefore is least susceptible to artifacts associated with heating, spatially nonuniform magnetic dynamics, and changes in the DC resistance in response to spin torque.[13,18] However, time-domain ST-FMR requires expensive, specialized equipment (*i.e.*, a high-bandwidth oscilloscope and multiple pulse generators). Here we show that it is possible to use a simple network analyzer measurement to determine the bias dependence of the spin torque vector, by studying the resonant response of a magnetic tunnel junction subject to both DC and microwave currents. We find excellent agreement with time-domain ST-FMR measurements[18] made on the same devices.

The MgO-based magnetic tunnel junctions (MTJs) that we study came from the same batches measured in references [18] and [19], with resistance-area products for the tunnel barriers equal to $RA$ = 1.5 $\Omega \cdot \mu m^2$ and 1.0 $\Omega \cdot \mu m^2$. We will present data for one sample with $RA$ = 1.5 $\Omega \cdot \mu m^2$, a resistance of 272 $\Omega$ in the parallel state, and a tunneling magnetoresistance (TMR) of 91%, but we found similar behavior in three other samples. The device on which we will focus has the layer structure (in nm): bottom electrode, IrMn pinned synthetic antiferromagnet (SAF) [IrMn(6.1)/CoFe(1.8)/Ru/CoFeB(2.0)], tunnel barrier [$MgO_x$], magnetic free layer [CoFe(0.5)/CoFeB(3.4)], capping layer [Ru(6.0)/Ta(3.0)/Ru(4.0)]. Both the pinned layer and the free layer were patterned into a circular cross section with a nominal 90 nm diameter. All the measurements were done at room temperature. We confirmed that the device properties did not degrade during the process of measurement[20] by checking that the device resistance and TMR remained unchanged. We will use a sign convention that positive values of current correspond to electron flow from the free layer to the reference layer (giving spin torque favoring antiparallel alignment).

We performed measurements with a commercial network analyzer (Agilent 8722ES, 50 MHz – 40 GHz) using the circuit in Fig. 1. We measured the microwave response in a reflection geometry, using a bias tee to allow simultaneous application of a DC bias to the MTJ. Before routing the reflected microwave signal to the network analyzer, we amplified it using a 15-dB amplifier in combination with a directional coupler. The microwave gain of the amplifier and transmission losses in other circuit components were calibrated by standard methods. Figure 2 shows an example of the real and imaginary parts of the reflected signal as a function of frequency, in the frequency range exhibiting spin-torque-driven magnetic resonance. These data correspond to a DC current of -0.4 mA and an applied magnetic field $H$ = 200 Oe oriented 70 degrees from the exchange bias of the SAF reference layer, so that the initial offset angle of the two magnetic layers is approximately $\theta$ = 61°. The microwave excitation signal $V_{in}$ that we applied to the sample had an amplitude always less than 22 mV. Within the model discussed below this results in magnetic precession angles < 3°, and we verified that the output signals scaled linearly with $V_{in}$ as expected in the linear-response regime.

To interpret these data, and to use them to measure the strength of the spin transfer torque, we analyze the reflected microwave signal $V_{ref}$ within a macrospin model of the magnetic dynamics, combining the Landau-Lifshitz-Gilbert-Slonczewski equation of motion for a magnetic tunnel junction subject to an oscillating spin torque together with appropriate microwave circuit equations. (See ref. [19] for details.) The resulting (complex-valued) reflection coefficient corresponding to the resonant magnetic response can be written

$$S_{11} \equiv \frac{V_{ref}}{V_{in}} = \frac{R_0 - (50\ \Omega)}{R_0 + (50\ \Omega)} + \frac{(50\ \Omega)}{R_0 + (50\ \Omega)} I_{DC} \chi(\omega), \tag{1}$$

where

$$\chi(\omega) \equiv \Delta R(\omega)/V_{in}$$

$$= -\frac{\partial R}{\partial \theta}\bigg|_I \frac{R_0}{R_0 + (50\,\Omega)} \frac{\gamma}{M_S Vol} \frac{1}{\omega - \omega_m - i\sigma} \left[ i\frac{\partial \tau_\parallel}{\partial V}\bigg|_\theta + \frac{\gamma N_x M_{eff}}{\omega_m} \frac{\partial \tau_\perp}{\partial V}\bigg|_\theta \right], \quad (2)$$

and the resonance frequency and current-dependent resonant linewidth are

$$\omega_m \approx \gamma M_{eff} \sqrt{N_x \left[ N_y - \frac{1}{M_{eff} M_S Vol} \left( \frac{\partial \tau_\perp}{\partial \theta}\bigg|_V + \frac{(50\,\Omega)}{R_0 + (50\,\Omega)} I_{DC} \frac{\partial R}{\partial \theta}\bigg|_I \frac{\partial \tau_\perp}{\partial V}\bigg|_\theta \right) \right]}, \quad (3)$$

$$\sigma \approx \frac{\alpha \gamma M_{eff}(N_x + N_y)}{2} - \frac{\gamma}{M_s Vol} \left( \frac{\partial \tau_\parallel}{\partial \theta}\bigg|_V + \frac{1}{2} \frac{(50\,\Omega)}{R_0 + (50\,\Omega)} I_{DC} \frac{\partial R}{\partial \theta}\bigg|_I \frac{\partial \tau_\parallel}{\partial V}\bigg|_\theta \right). \quad (4)$$

Here $R_0$ is the differential resistance of the MTJ, $\Delta R(\omega)$ is the oscillating part of the DC resistance, $\theta$ is the angle between the magnetizations of the two electrodes of the MTJ, $\alpha$ is the Gilbert damping parameter, $M_S Vol$ is the total magnetic moment of the free layer, $\tau_\parallel(V,\theta)$ and $\tau_\perp(V,\theta)$ are the "in-plane" and "perpendicular" components of the spin torque, $V$ is the voltage across the MTJ including both DC and high-frequency terms, $\gamma = 2\mu_B/\hbar$ is the absolute value of the gyromagnetic ratio, $N_x = 4\pi + H/M_{eff}$, $N_y \approx H/M_{eff}$, $4\pi M_{eff}$ is the strength of the easy-plane anisotropy field, and $H$ is the component of applied magnetic field along the precession axis. When both in-plane and perpendicular components of torque are present, both the real and imaginary parts of the resonant signal consist of a sum of frequency-symmetric and antisymmetric Lorentzian curves. Both torque components can therefore be extracted by fitting the symmetric and antisymmetric parts of either the real or imaginary response.

The solid lines in Fig. 2(a) and (b) show an example of the good agreement we find when fitting Eq. (1) to our resonance measurements. We observe two resonances in each panel in Fig. 2, one with large amplitude near 5.9 GHz and a second with smaller amplitude near 7.5 GHz. We perform separate fits to the real and imaginary curves, employing four free parameters for each resonance in a fit: the center frequency of the resonance, the amplitude of the frequency-symmetric and antisymmetric Lorentzians, and the linewidth (taken to be the same for both the symmetric and antisymmetric components). We allow for a small nonzero constant slope in the non-resonant background signals (dashed lines in Fig. 2) that may be associated with an imperfect capacitance calibration.

The dependences on $H$ and $I_{DC}$ for the real part of the resonances are shown in Fig. 3(a) and 3(b). As in Fig. 2(a), the spectra contain one primary dip in Re($S_{11}$) together with a smaller side resonance at a higher frequency. The primary resonance shifts with $H$ as expected from the Kittel formula while the secondary signal shifts more slowly and decreases in amplitude with increasing field strength. We suspect that the secondary peak may involve coupled motion of the magnetic layers in the synthetic antiferromagnet polarizing layer. To avoid having this mode interfere with our measurements of spin torque, we select values of magnetic field and magnetic field angle such that the secondary mode has small amplitude and maximum separation in frequency from the primary mode. These are the same selection criteria used in ref. [18].

Based on Equations (1) and (2), for any value of bias we can determine the spin transfer "torkances"[21] $\partial \tau_{\|} / \partial V \big|_\theta$ and $\partial \tau_{\perp} / \partial V \big|_\theta$ from fits to the frequency-symmetric and antisymmetric parts of the primary resonance in either Re($S_{11}$) or Im($S_{11}$). In calculating the torkances from the resonant amplitudes we use the following parameters:

$M_S Vol = 1.8 \times 10^{-14}$ emu ($\pm 15\%$),[18] $4\pi M_{eff} = 13 \pm 1$ kOe determined from high-field measurements of the resonance frequency, and $\alpha = 0.016 \pm 0.001$ determined by measuring the resonance linewidth at positive and negative biases and interpolating to zero bias. In Fig. 3(c) we plot the bias dependence of the resulting torkances as found by the network analyzer technique. We normalize the results by $\sin\theta$ since the spin torque of a MTJ is predicted to have this angular dependence.[21] We note that the torkance values determined by independent fits to the real and imaginary parts of the resonance agree, as is required in order that our analysis procedure be self-consistent. Figure 3(c) also shows a comparison to measurements on the same sample using the time-domain ST-FMR technique introduced in ref. [18], whereby the magnetic precession driven by a resonant spin torque is detected by a fast oscilloscope. We find excellent agreement between the two types of measurements. The in-plane component of the torkance, $\partial \tau_\parallel / \partial V |_\theta$, measured by the two techniques agrees in magnitude near zero bias with the same moderate dependence on bias, with no adjustment of parameters for either technique. The perpendicular component $\partial \tau_\perp / \partial V |_\theta$ displays the same approximately linear bias dependence at low bias. In Fig. 3(d), we plot the full bias dependent torques $\tau_\parallel(V)$ and $\tau_\perp(V)$, obtained by numerical integration of the torkances.

Neither the network-analyzer ST-FMR technique nor the time-domain ST-FMR technique can be used at $V = 0$, because a non-zero DC bias is required to generate the oscillatory voltage signal that is measured (see Fig. 3(c)). (For measurements near zero bias, DC-voltage-detected ST-FMR can provide accurate torque measurements without artifacts in the mixing signal.[11-13]) The time-domain ST-FMR technique allows measurements to higher biases, because it is naturally implemented using short bias pulses that are less likely to produce dielectric breakdown in the tunnel barrier, compared to the constant DC biases used in our network analyzer technique. However, in the bias range shown in Fig. 3(c,d) the network analyzer method provides a more convenient approach in that it does not require specialized, expensive equipment, while it yields a sensitivity comparable to time-domain ST-FMR.

In summary, we demonstrate that it is possible to use a simple network-analyzer technique to measure the strength and direction of the spin transfer torque vector as a function of bias in magnetic tunnel junctions. This technique provides roughly similar sensitivity as the time-domain ST-FMR method,[18] making it useful as a simple and rapid means for characterizing spin-torque devices.

Cornell acknowledges support from ARO, NSF (DMR-1010768), ONR and the NSF/NSEC program through the Cornell Center for Nanoscale Systems. We also acknowledge NSF support through use of the Cornell Nanofabrication Facility/NNIN and the Cornell Center for Materials Research facilities (DMR-1120296).

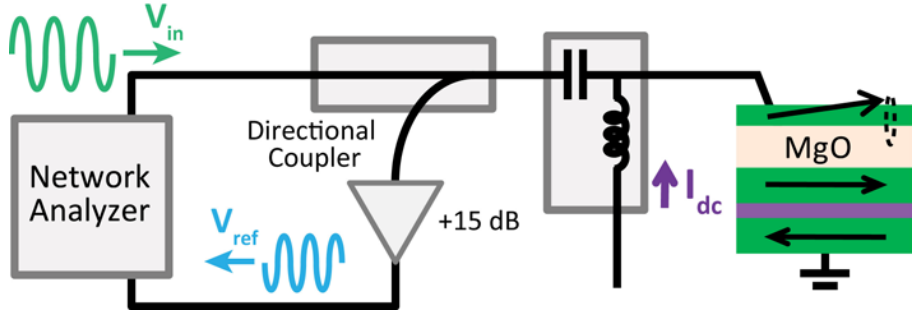

FIG. 1. (color online) The network analyzer circuit used in the measurement.

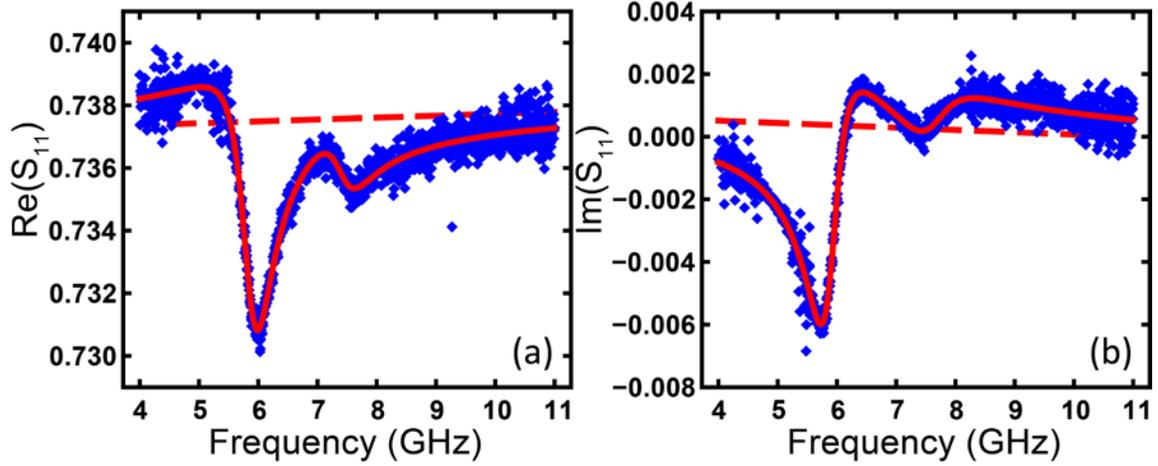

FIG. 2. (color online) The measured (a) real part and (b) imaginary part of the reflection signal ($S_{11}$) for $I_{DC}$ = -0.4 mA and a magnetic field $H$ = 200 Oe applied 70° from the exchange bias direction, giving $\theta$ = 61°. The solid lines are a fit to Eq. (1). The dashed lines are the nonresonant backgrounds used in the fits.

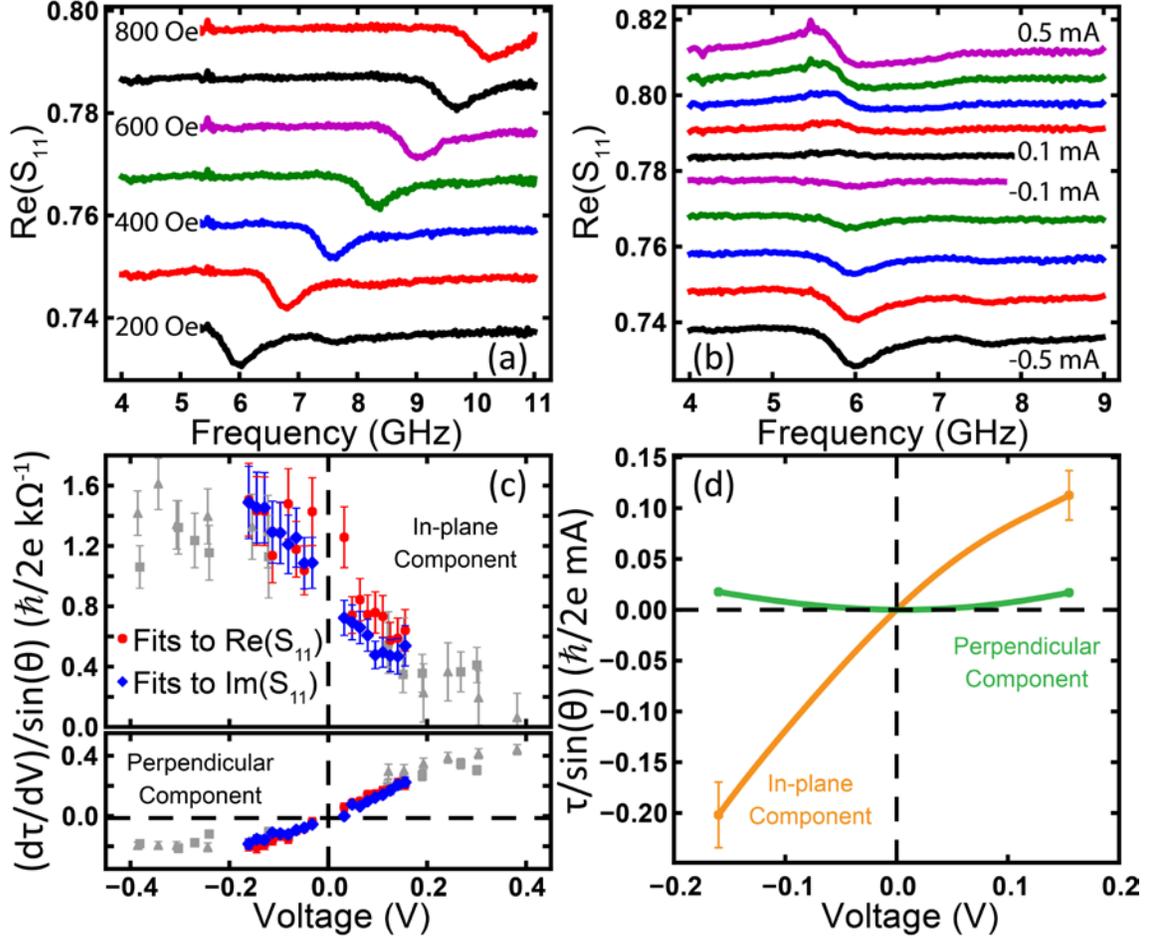

FIG. 3. (color online) (a) Measured frequency dependence of the real part of $S_{11}$ for several values of magnetic field applied 70° from the exchange bias direction, with $I_{DC}$ = -0.4 mA. The curves are offset by 0.01 vertically. (b) Measured frequency dependence of the real part of $S_{11}$ for several values of DC current, with $H$ = 200 Oe applied 70° from the exchange bias direction. The curves are offset vertically by 0.01. (c) Bias dependence of the in-plane and perpendicular components of the torkance $\partial \tau / \partial V |_\theta$ determined by fitting to the frequency dependence of Re($S_{11}$) (red circles) and Im($S_{11}$) (blue diamonds) at different values of the DC bias. These data correspond to $H$ = 200 Oe applied 70° from the exchange bias direction, giving $\theta$ = 61°. For comparison we also show in gray

the results on the same device from time-domain ST-FMR measurements (triangles: for $H$ = 250 Oe applied 95° from the exchange bias direction giving $\theta = 85°$; squares: $H = 200$ Oe applied at 68° giving $\theta = 64°$). (d) Integrated in-plane and perpendicular components of the spin torque vector determined by integrating the network-analyzer data in (c), with representative error bars.